\documentclass[prl,twocolumn,showpacs]{revtex4} 
\newcommand{\pad}[2]{\frac{\partial #1}{\partial #2}}
\usepackage{graphicx}
\begin{document}

\title{A Hybrid Monte Carlo Method for Surface Growth Simulations}
\author{G. Russo}
\affiliation{Dipartimento di Matematica ed Informatica, 
Universit\`a di Catania,  95125 Catania, Italy}
\author{L. M. Sander}
\affiliation{Michigan Center for Theoretical Physics and Department of Physics,
	University of Michigan, Ann Arbor, Michigan, 48109-1120}
\author{P. Smereka}
\affiliation{Michigan Center for Theoretical Physics and Department of Mathematics,
	University of Michigan, Ann Arbor, Michigan, 48109-1120}
\date{\today}

\begin{abstract}
We introduce an algorithm for treating growth on surfaces which combines important
features of continuum methods (such as the level-set method) and Kinetic Monte Carlo (KMC) simulations. 
We treat the motion of adatoms in continuum theory, but attach them to islands one atom at a time. The technique is borrowed from the Dielectric Breakdown Model. Our method allows us to give a realistic account of 
fluctuations in island shape, which is lacking in deterministic continuum treatments and which is an important physical effect. Our method should be most important for problems close to equilibrium where KMC becomes 
impractically slow. 
\end{abstract}
\maketitle 
Epitaxial growth on surfaces is of central importance both for applications and as a very interesting example of statistical processes out of equilibrium. We may idealize the process as the introduction of new atoms (adatoms) onto a crystal surface with flux, $F$; the adatoms then diffuse, with diffusion coefficient, $D$, nucleate islands or attach to existing islands. During the early stages of growth, the submonolayer case, the island size and shape distribution is a matter of substantial practical and theoretical interest.

The island growth process is commonly modeled by
kinetic Monte Carlo (KMC) or continuum models. In KMC, internal noise
processes are automatically represented within the model and each adatom
is represented individually. Therefore, when there are many adatoms
(e.g. close to equilbrium) such simulations slow down considerably. A
deterministic continuum model which represents the adatoms as a
continuous fluid does not have this problem, and should be much faster.
There has been considerable work in the development of such  models for
epitaxial growth (see \cite{gyure98,chen01} and references therein).  In some cases
they have been quite successful, but sometimes they do not reproduce
experimental results. One reason for such problems is that
deterministic continuum models neglect important fluctuations. In this
paper we present a method of dealing with some fluctuations without giving up
the advantages of a continuum treatment. We call this approach Hybrid 
Monte Carlo (HMC). The most important use of this method 
will be in cases where fluctuations are important, but
which would be difficult to treat with KMC because of the presence of a large number
of adatoms. A computation in a similar spirit has been given in \cite{schulze01,schulze03}. However, the present approach differs in a number of important ways.

There are several sources of fluctuations in the growth process. The one we consider 
here is the fact that when atoms attach to islands they do so one at a time
 -- that is, there is {\em shot noise} in the island growth process.  
This is important because island growth limited by diffusion is {\em intrinsically unstable}. 
The surface of the island will grow fingers due to the
analogue of the well- known Mullins-Sekerka instability of metallurgy
\cite{mullins63}.  In the context of thin film growth the instability
was discussed in detail by Bales and Zangwill \cite{bales90}.  The
reason for unstable growth is easy to see: if a finger on the edge of an island
starts to grow it will project out onto the terrace and be fed by more
adatoms than the portions behind.  The finger will grow longer, and be
fed by still more flux, etc.  Edge diffusion and other restructuring
processes smooth out the fingers, and the final shape depends on the
competition between unstable growth and smoothing.

An extreme case of the diffusively unstable
growth is represented by the Diffusion-Limited Aggregation (DLA) model of Witten and
Sander \cite{witten81}.  In this model all smoothing processes are neglected
and growth takes place so slowly that one random walking adatom at a
time is considered. DLA clusters are sprawling fractal objects with many 
branches which resemble some cases of island growth \cite{hwang91,tsui96}.
There is a variation of the model called the 
Dielectric Breakdown Model (DBM) \cite{niemeyer84} in which random walkers 
are not used. Instead the Laplace equation is solved outside the aggregate for a field, $\rho$,
which represents the probability density of walkers, and the growth
algorithm is to add one particle at a time with probability
proportional to $\partial \rho/\partial n$ at the surface. Thus the DLA limit
can be successfully treated by a model in which the adatoms are a continuum. However,
methods such as the level-set method \cite{chen01} cannot go to this limit and thus 
cannot produce dendritic islands which are seen in experiment.

The unstable modes for the interface of the islands are present in level-set models, 
of course. However,  the reason why the DBM limit is not achieved is
that the growth process is represented by
deterministically advancing a continuum interface
according to the flux of the adatom fluid into the surface (see Eq.~(\ref{speed}) below). 
In this algorithm the amplitude of the perturbations to a smooth interface 
are not correctly represented: they are given either by the initial conditions or by computer 
roundoff errors. In the experiment, however, there is a mechanism for feeding  the instability: 
each adatom attaches not as a spread-out advance of the interface, but as an atom. In the case of DLA the result is that noise in the shape is present at all scales \cite{somfai99} and does not average out.

We should mention that there are other fluctuations which we will not treat. For example, there are
density fluctuations in the adatom fluid which are important in the nucleation of new islands. There is a
method \cite{ratsch02} to treat this within level-set theory that we could employ. We
will not consider such processes in this work, but rather look at the shape fluctuations of existing islands.

The HMC algorithm goes as follows: we treat the islands on the surface as crystals 
containing discrete atoms which occupy the sites of a lattice. To illustrate the 
method we use a square lattice here. On the other hand, 
the adatoms are treated as a continuum whose surface density is $\rho$, and which is 
governed by:
\begin{equation}  
\partial_t \rho = D\nabla^2\rho+F
\label{level}  
\end{equation}
In practice we solve this equation numerically on a discrete square grid which is
commensurate with the crystal. 

Eq. (\ref{level}) is solved with periodic boundary conditions 
on the edge of the system. The important physics
of growth in incorporated into the boundary condition at
the surface of the island. We put:
\begin{equation}  
D\pad{\rho}{n}=\mp k_{\pm}(\rho - \rho_o)  
\label{bc}
\end{equation}  
Here $k_-$ and $k_+$ are the respective attachment rates   
on the upper and lower  edge of the island boundary. 

In the case of irreversible growth only the first term in brackets in
Eq. (\ref{bc}) would be present. The other term,  
 $\rho_o$,  accounts for the
detachment of atoms from the island, and depends on the
position on the island boundary. Physically, the boundary condition
must allow for faster detachment at corners, say, than at flat surfaces. 
We represent this in a way that allows us to compare directly with the bond-counting
version of KMC:
\begin{equation}
\rho_o =  \exp(-nE/k_{B}T)
\label{rhoeq}
\end{equation}
Here $n$ is the number of nearest neighbor bonds that must be broken to 
\emph{completely detach} (see below) the atom
in question and add it to the adatom sea. The value of $n$ depends on the environment
of the detaching atom.
 We imagine that atoms can break
bonds by moving along lattice directions.  Also,
$E$ is the bond energy, and $T$ the temperature. Of course, we 
can easily incorporate bonding to more distant neighbors.

In a pure continuum model, the velocity of the island growth
would be determined by mass conservation:
\begin{equation}  
v_n=a^2D\left[\pad{\rho}{n}\right]
\label{speed}  
\end{equation}  
where $a$ is the lattice constant and $\left[\cdot\right]$ denotes the
jump across the island boundary.  
Note that we can interpret Eq. (\ref{bc}) in a way which is familiar
in studies of crystal growth \cite{langer80}. By combining Eqs. (\ref{bc},\ref{speed}) we find 
an equation for the density at the surface:
\begin{equation}
\rho - \overline{\rho} = (\rho_o- \overline{\rho}) + \alpha v_n
\label{dendritic}
\end{equation}
where $\alpha=1/(a^2[k_+ +k_-])$, and $\overline{\rho}$ is the
equilibrium density near a flat surface. That is, we are including both local equilibrium
and kinetic terms in our boundary condition. The difference $\rho_o - \overline{\rho}$ 
is a measure of the number of dangling bonds on the surface, and thus of the curvature (upon
 coarse-graining). That is, the first term in Eq. (\ref{dendritic}) is related to the familiar 
Gibbs-Thompson boundary condition of crystal growth, and the second is a kinetic term.

In HMC we implement Eq. (\ref{speed}) in a  way
that \emph{includes fluctuations}. Consider first a
case where attachment is the only important process. Then  
 we solve Eq.~(\ref{level}, \ref{bc}) and compute the total flux  
onto the island boundary using Eq. (\ref{bc}, \ref{speed}). When the total flux exceeds one atom  
then an adatom is attached  to the boundary at random with  
the probability proportional to $v_n$ (exactly as in the DBM model.) In the
case where detachment is also present 
we consider the surface to be partitioned into the part where the net flux is
inward (growth), and attach atoms with probability density $\propto v_n$, and outward 
(detachment) and remove island atoms with probability density $\propto -v_n$.

We have implemented HMC and compared the results to a KMC code on the same square lattice using
nearest neighbor bonding. We use the 
hopping rate of an adatom to set the unit of time,  and the lattice constant to be unity.
We have two independent parameters 
namely $D/F$, and $\epsilon=E/k_BT$ 
We can also add edge diffusion, but here we have not done so.
The KMC code is written using the method described in \cite{newman99} which takes advantage of the fact that there are
only a few independent jump probabilities for a bond-counting model. 
We find that, as expected, at large $\epsilon$, the KMC code is much faster. However, for 
$\epsilon=1.5$ and with about 300 adatoms in a 40x40 system the speeds are comparable. 
We note that in situations such as heteroepitaxy
\cite{lam02} there are a great many independent probabilities to jump and KMC is slower.
 
\begin{figure}
\parbox{1.5in}
{\includegraphics[width=1.4in]{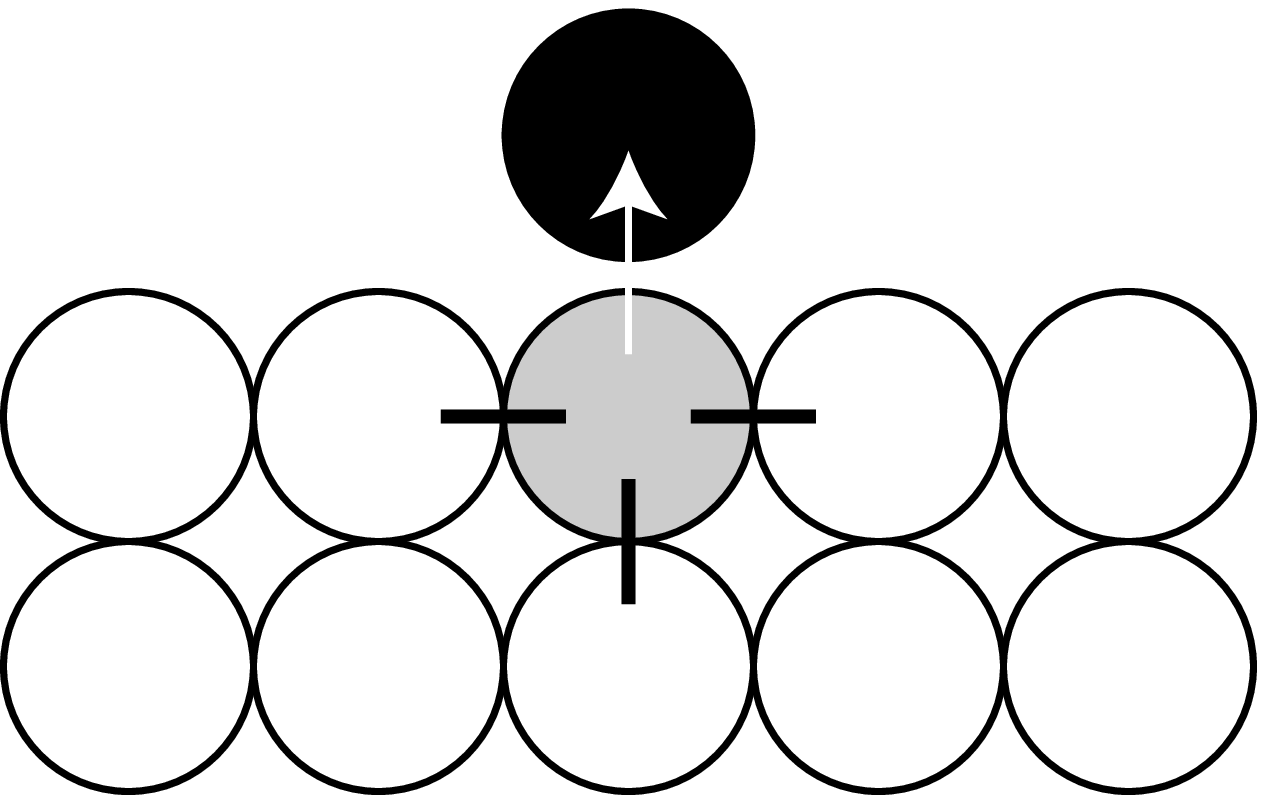}}
\hspace{.1in}
\parbox{1.5in}
{\includegraphics[width=1.1in]{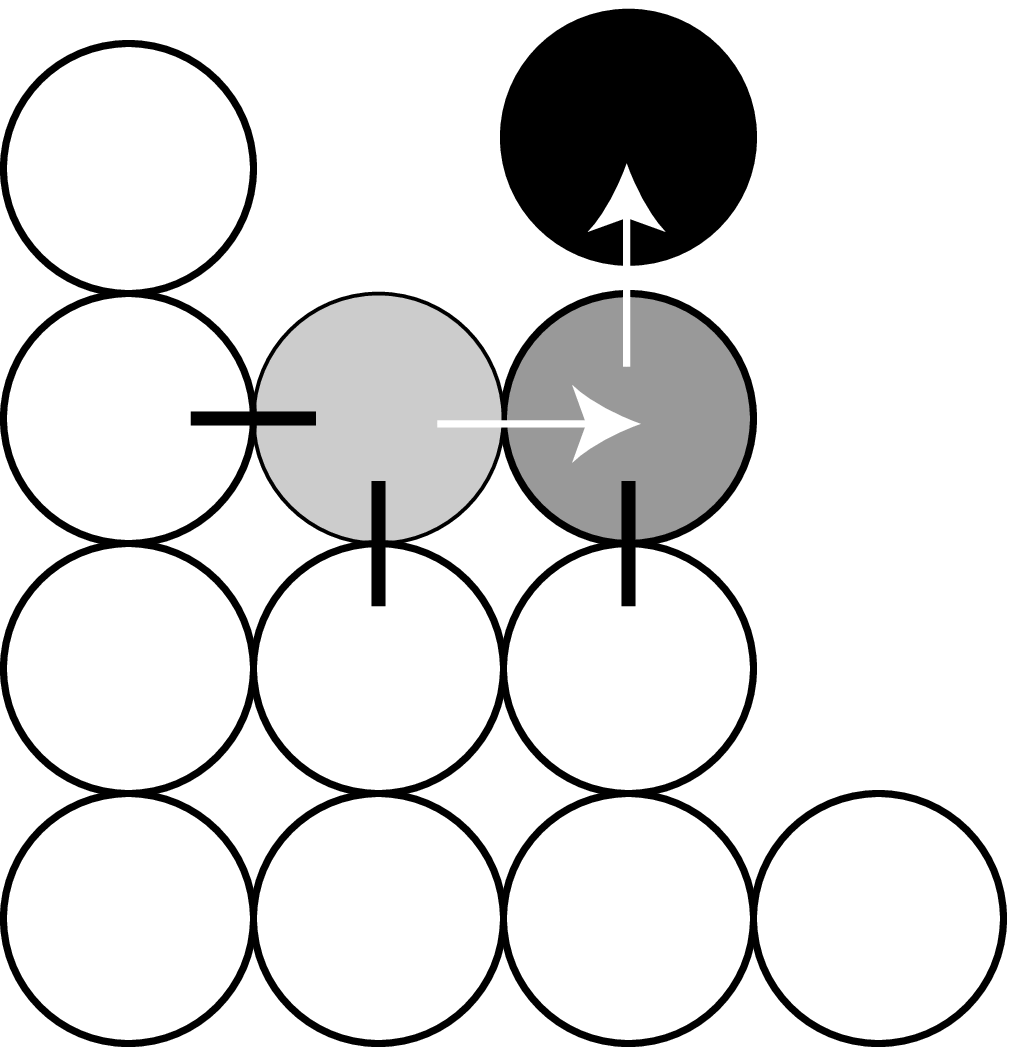}}
\caption{a.) Detaching from a $[10]$ surface can be done in one step, and breaks 3 bonds. 
b.)Fully detaching from a $[11]$ surface in two steps also breaks 3 bonds.} 
\label{sketch}
\end{figure}

The interpretation of the number of bonds, $n$, is a bit delicate. In Fig. (\ref{sketch}) we
show some examples of what we mean by `complete detachment'.
For our square lattice, to detach an atom on a  $[10]$ surface 
we need to break 3 bonds. However, for a $[11]$ surface, in order to detach from the surface,
an atom must first break two bonds, and then subsequently, one more --  see Fig. (\ref{sketch}).
Thus, for a $[11]$ surface we set $n=3$ as well because this corresponds to  the product of the probabilities of the two processes. In effect, we have coarse-grained. For other possible surface environments it is not difficult to tabulate the
correct value of $n$. 

\begin{figure}
\parbox{1.4in}
{\includegraphics[width=1.4in]{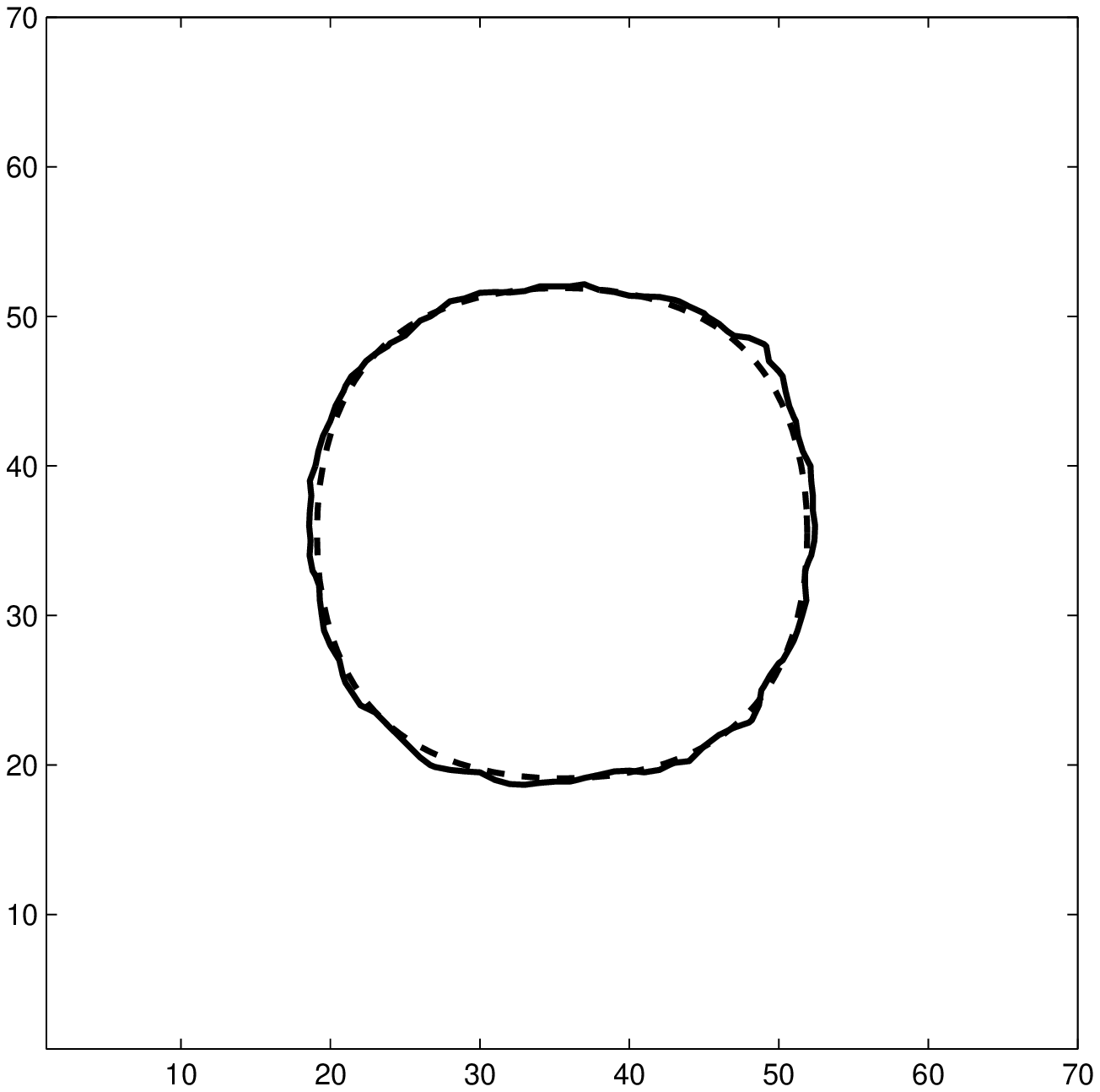}}
\hspace{.1in}
\parbox{1.4in}
{\includegraphics[width=1.4in]{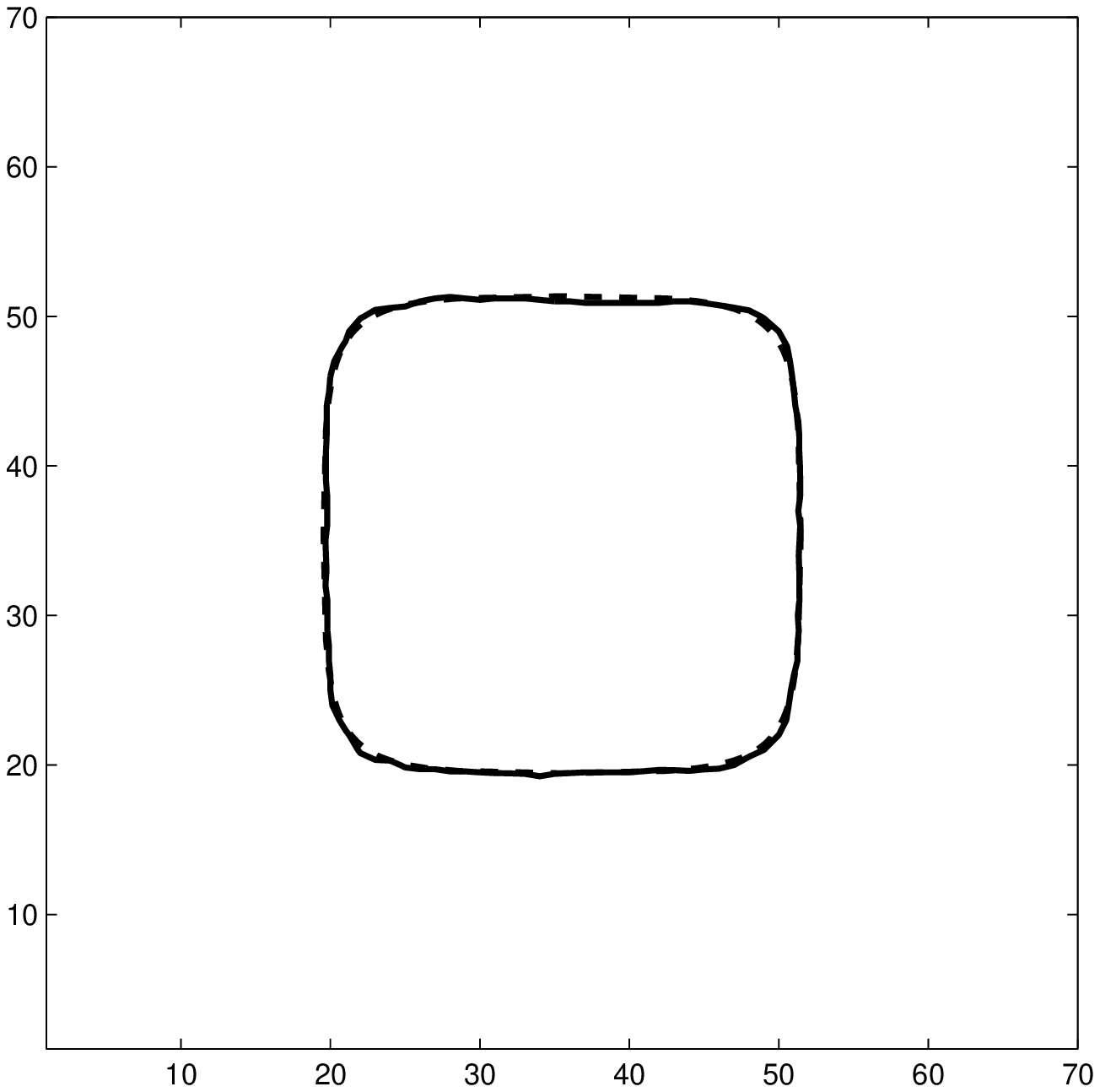}}
\caption{HMC simulations of the average profile of an equilibrium island compared with
exact results. Left is for $\epsilon=2$ and right for $\epsilon=5$. The heavy line is the exact result
and the dotted line from the simulation. The agreement is excellent.}
\label{equilshape}
\end{figure}

This procedure may seem arbitrary, but we can justify it by its results. To this end, we show that
HMC gives the correct shape for an island in equilibrium with adatoms. This shape is known exactly \cite{rottman81} from a mapping to the 2d Ising model.   We did simulations  with $F=0$ starting with a square island,
and ran our code for a long enough time that the system seemed to be in equilibrium.  
In Fig. (\ref{equilshape}) we show some results
superimposed on the exact results for various temperatures. The 
simulation results are ensemble averages; that is, we did 20 independent simulations and 
averaged the density of the island after shifting the center of mass of each one to be at the origin. 
The dotted line in Figure \ref{equilshape} is the contour line 
where the averaged island density is 1/2.

\begin{figure}
\includegraphics*[width=2.1in]{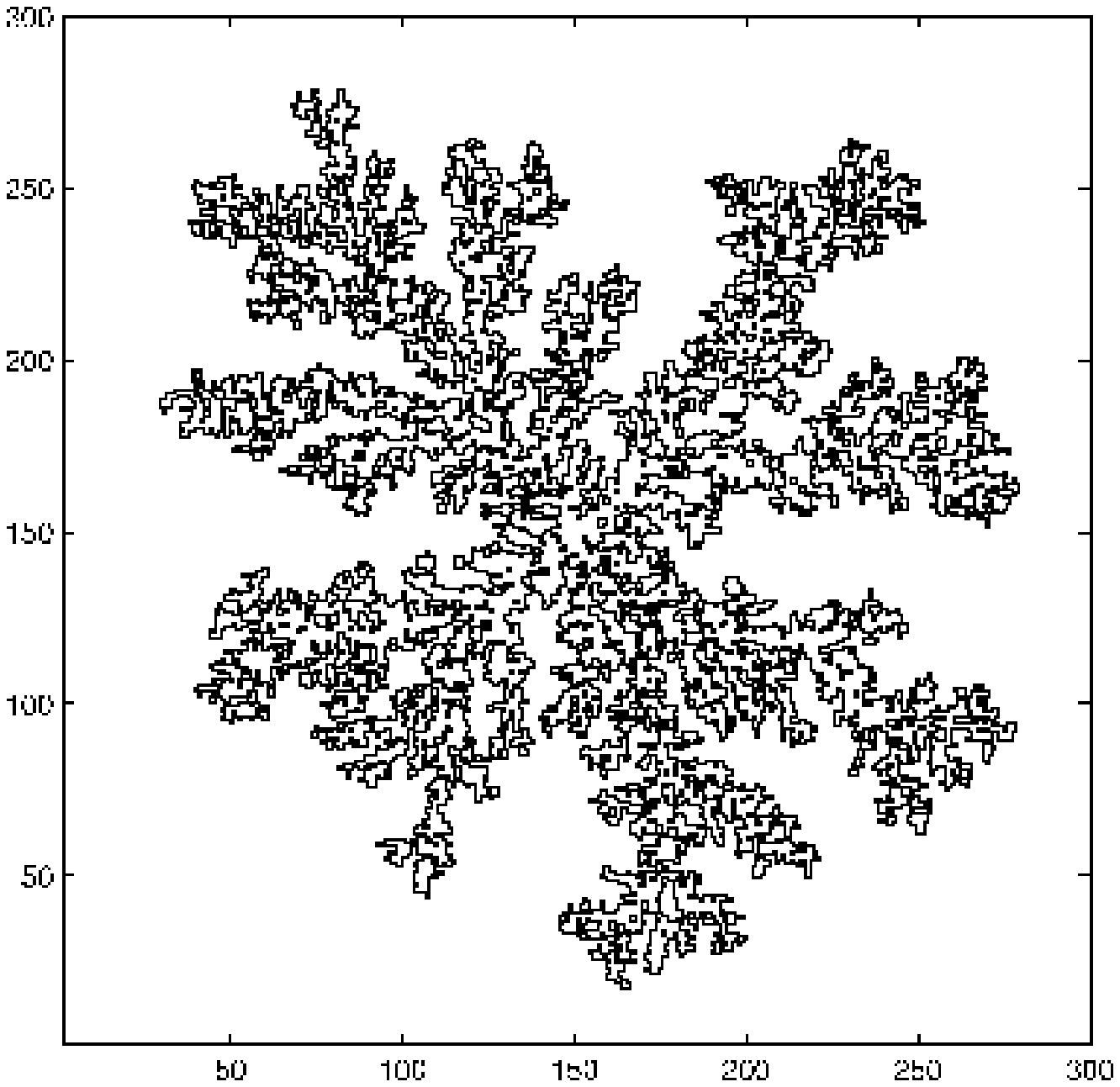}
\includegraphics[width=2.1in]{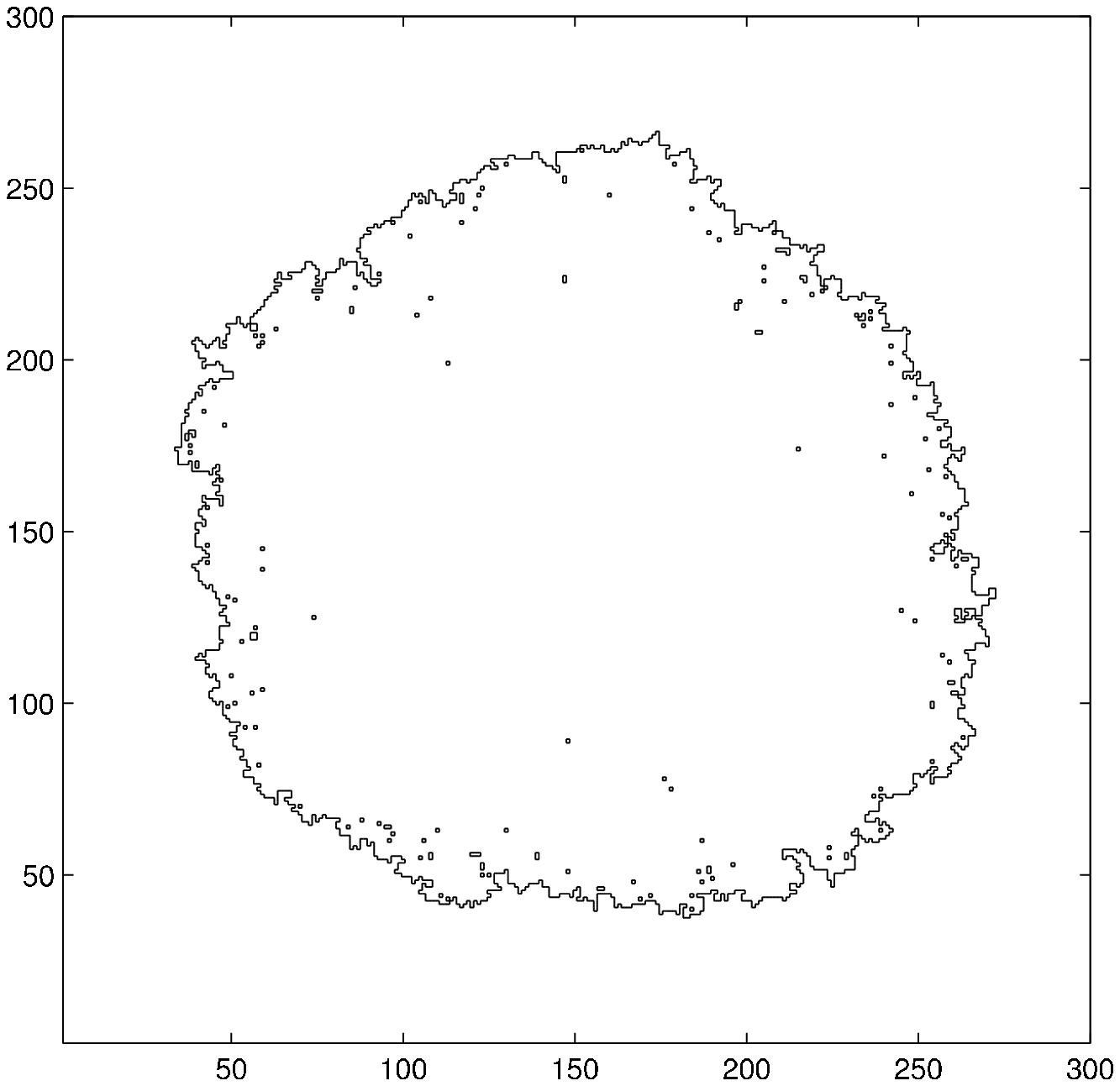}
\caption{Irreversible (low $T$) growth of islands for two values of    $D/F$.
Top,  $D/F=10^5$. Bottom, $D/F=10^2$. }
\label{DLA1}
\end{figure}

\begin{figure}
\includegraphics[width=2.1in]{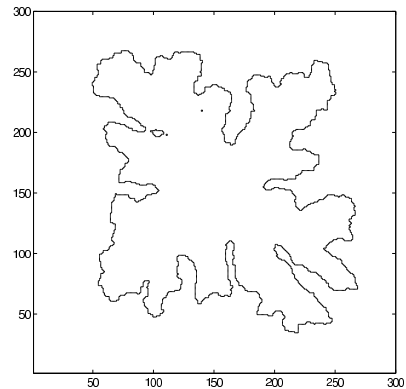}
\caption{Island growth with $D/F=10^5$ and $\epsilon=2.5$}
\label{DLA2}
\end{figure}

We now present some more results to demonstrate the technique.  
If growth dominates detachment we should have DLA-like structures. Whether this occurs  depends on $T$
and  $D/F$ \cite{barabasi95}. For low temperature, Fig. (\ref{DLA1}), we see the transition in a very clear way. Fig. (\ref{DLA2}) shows a higher $T$ case.
Of course, all of these effects can be seen in KMC simulations. 
In other simulations (not shown), we have demonstrated that edge diffusion also smoothes out dendritic shapes,
as expected.
 The virtue of our method will be to
treat systems near equilibrium where the dynamics of fluctuations are of interest and where there are a great number
of adatoms,  as in Fig. (\ref{DLA2}). 

\begin{figure}
\includegraphics[width=2.6in]{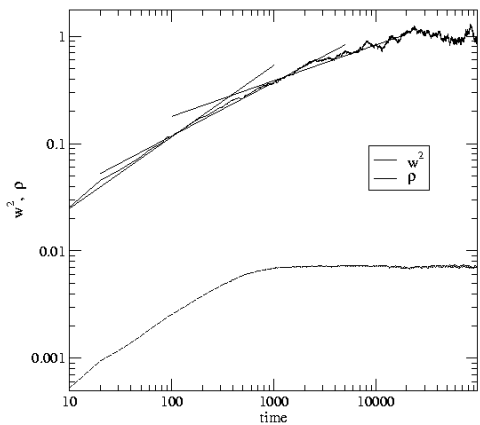}
\caption{Thermal roughening of  a pair of steps, average over 20 simulations. Lower curve (dotted) is the average adatom density, $\rho$. For early times, while the steps are moving, and $\rho$ is changing,   the scaling is $w^2 \propto t^{2/3}$, kinetic
roughening.  Later there is thermal roughening with  $w^2 \propto t^{1/2}$ (evaporation-condensation kinetics) crossing over to terrace-diffusion kinetics, $w^2 \propto t^{1/3}$. The straight lines, from left to right,  have slopes 2/3, 1/2, and 1/3. } 
\label{step}
\end{figure}

Another example is the thermal broadening of steps due to repeated attachment and detachment of adatoms.
There is an extensive literature in this area. \cite{jeong99}, and the theoretical expectation
\cite{jeong99,khare98} is that the thermal width, $w$, of a step should depend on the rate-limiting mechanism for
step motion. In our model, without surface diffusion, this will be
the either detachment from the step or diffusion on the terrace. In these cases $w^2  \propto t^{1/2}, t^{1/3}$, respectively.   
We show, in Fig. (\ref{step}),  $w^2(t)$  for a pair of  steps one atomic layer high which cross 
a $100 \times 100$ terrace. We see indications of
both of the expected behaviors at late times. The early time behavior is \emph{kinetic} roughening \cite{kardar86}, $w^2 \propto t^{2/3}$. This is because  we started with no adatoms, and, initially,   the steps were retreating.

The HMC technique should be most useful in situations where there is a large separation of time scales between the diffusion of the adatoms and fluctuations of island boundaries. Other examples for which it might be used are in studies of homogeneous nucleation of large islands near equilibrium on surfaces \cite{theis96}. Another example is in the catalysis of the reaction CO+O 
$\rightarrow$ CO$_2$ on a noble metal. The diffusion of CO is very fast in this case, and makes a realistic KMC simulation
very difficult. Evans and collaborators \cite{suchorski01} have used an approximation where $\rho_{CO}$ is constant where ever
there is no O on the surface. Our approach (with suitable changes in boundary conditions to allow for the reaction) should
give a better account of the kinetics. 

Finally, we should mention the case of heteroepitaxy. There have been a number of KMC simulations of this
very important process \cite{lam02,orr92,khor00}. This is a very difficult problem to attack numerically. 
The time-consuming step in such calculations is not the dynamics of the adatoms, but rather the elasticity computation. Nevertheless, our approach to the adatoms offers a number of advantages, and we intend to pursue this in a future
publication.

We acknowledge useful conversations with B. Orr, R. Calflisch, and J. Evans. PS is supported by NSF grant DMS 0207402. GR acknowledges a travel grant from the Michigan Center for Theoretical Physics.

\bibliography{hmc3}







\end{document}